\documentclass{ws-procs975x65}

\begin{document}

\title{Just enough inflation}

\author{D.J.SCHWARZ\footnote{Talk presented at MG12, Paris, 2009} and E.RAMIREZ}

\address{Fakult\"at f\"ur Physik, Universit\"at Bielefeld,\\
Postfach 100151, 33501 Bielefeld, Germany\\
E-mail: dschwarz, eramirez  at physik.uni-bielefeld.de}

\begin{abstract}
We propose a version of chaotic inflation, in which a fundamental scale $M$, well below the 
Planck scale $M_{\rm P}$, fixes the initial value of the effective potential. If this scale happens to be 
the scale of grand unified theories, there are just enough e-foldings of inflation.  An initial epoch 
of fast-roll breaks scale-invariance at the largest observable scales.   
\end{abstract}


\bodymatter

\section{Motivation}

Cosmological inflation provides an explanation for the formation of structures 
in the Universe. However, we lack a fundamental theory from 
which inflation would emerge as a natural consequence.
Instead, a large number of inflationary scenarios have been proposed. 
On the other hand, the standard model of particle physics (SM) unifies 
three of the four fundamental forces of nature and requires the existence 
of a particle with the same properties as the field driving inflation in many models. 

If the Higgs field could act as the inflaton field, current particle physics experiments at Tevatron 
and LHC would probe the physics of inflation. The study of this problem has first been 
addressed long ago \cite{linde82}. It was shown 
that a false vacuum scenario for the Higgs field (at the electroweak energy scale) leads to an overproduction of density perturbations and is thus ruled out.  An alternative would be to 
concentrate on much higher values of the Higgs potential in the spirit of 
chaotic (or large field) inflation \cite{linde83}. 

This scenario has been investigated in the context of a Higgs field minimally 
coupled to gravity, approximating the potential by its quartic term \cite{irst} 
and in theories with a non-minimal coupling \cite{bscd}. 
It has been shown \cite{running} that the SM could in principle be well defined 
up to the Planck scale. Then the $\lambda\phi^4$ potential, presumably the dominant 
contribution to the effective Higgs potential at high energies, could provide enough inflation, 
but generically overproduces density perturbations as $\lambda = {\cal O}(1)$ 
and is thus ruled out. Even if one allows for an arbitrary value of $\lambda$, data from 
5 years of WMAP observations exclude the scenario of an extended slow-roll epoch in a 
$\lambda \phi^4$-potential \cite{wmap5}.   
Likewise, the inclusion of a non-minimal coupling allows the possibility of adjusting 
the Higgs potential so that cosmological perturbations can be in accordance 
with observational restrictions \cite{bscd}.

\section{Fundamental scales} 

Here we consider a different situation in which the total amount of inflation
is not much more than 60 e-foldings. The onset of inflation is 
thus observable and violates the slow-roll assumptions. The resulting primordial 
power spectrum is not scale invariant, as the moment of the onset of the 
slow-roll regime distinguishes a scale. It turns out that such a situation cannot be 
discarded on grounds of current analysis and observations\cite{RS}. 

This new scenario of ``just enough'' chaotic inflation seems to be generic 
if two fundamental energy scales are relevant in the very early Universe. The
Planck scale $M_p$ is the fundamental scale of quantum gravity. The notions 
of spatial curvature, expansion rate $H$ and kinetic energy density 
are well defined and real quantities, at least up to that scale. 
However, this is less clear for the effective potential $V$ of the inflaton. 
The effective potential carries the information about all interactions 
of the inflaton, except its gravitational ones. If we consider the SM Higgs as a
candidate for the inflaton, it is clear that inflation would only take place if the
effective Higgs potential is real --- the existance of an imaginary part would lead 
to the immediate decay of the inflaton.   

The quartic self-coupling of the Higgs runs with energy. Depending on the mass 
of the top-quark, the self-coupling decreases with increasing energy scale $\mu$ and can 
become negative at high energy (see Fig.~1). The effective Higgs potential 
is real as long as the quartic self-coupling is positive, but becomes complex 
as soon as it runs to negative values \cite{fjse}. Such an imaginary contribution 
would give rise to the decay of the Higgs field and thus establishes an effective 
upper bound for an effective real Higgs potential $V < M^4$, with $M < M_{\rm P}$.
Motivated by the SM Higgs, we can assume that $M \sim 10^{15}$ GeV and $\lambda \ll 1$ 
at that scale. 

\begin{figure}
\psfig{file=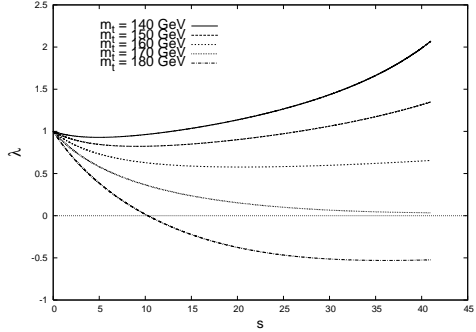,width=2.5in}
\caption{Running of $\lambda$ for various choices of the (bare) top quark mass 
as a function of $s \equiv \ln(\mu/m_Z)$ from $s=0$ ($\mu = m_Z$) to $s = 41$ 
($\mu \approx 24 M_p$). Here we assume $\lambda = 1$ and $\alpha_3=0.12$ 
at $s=0$ and use the renormalisation group equations from the SM at one loop.}
\label{lambda}
\end{figure}

\section{$\lambda \phi^4$-inflation}

We applied that idea to $\lambda \phi^4$-inflation \cite{RS}. As for chaotic inflation, 
curvature, expansion rate and kinetic energy density start at the Planck scale. However, if the 
potential energy were at $M_{\rm P}$ as well, the inflaton would decay and the Universe 
would not grow old. Only regions with $V < M^4$ (occurs with a finite probability) start to 
inflate. These regions are dominated by kinetic energy initially. It decays quickly 
and the Universe enters a state of fast-roll inflation, followed by about 60 e-foldings of slow-roll inflation. 
The trajectory of the Universe in the observable plane of tensor-to-scalar ratio vs. spectral tilt is 
shown in Fig.~2. It turns out that this model evades the constraints form the WMAP 5 year data
\cite{wmap5}, as it breaks scale-invariance at the largest cosmological scales \cite{RS}.
 
We found that by means of a fine tuned running of $\lambda$, the tree-level SM Higgs effective 
potential could give rise to such a scenario. However, the inclusion of one- and higher-loop 
corrections spoils that consistency. 

\begin{figure}
\psfig{file=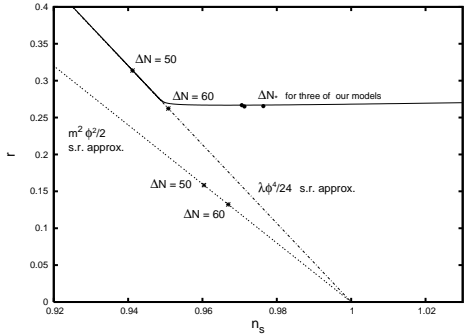,width=2.5in}
\caption{Tensor-to-scalar ratio $r$ versus the spectral index $n_{\rm s}$. The 
trajectory of just enough inflation violates the slow-roll approximation. 
Two standard slow-roll trajectories are shown for comparison.}
\label{observ_12}
\end{figure}

%\section{Acknowledgment}
\noindent 
We acknowledge financial support by the Deutsche Forschungsgemeinschaft (DFG).


\begin{thebibliography}{99}
\bibitem{linde82}
  A.~D.~Linde,
%  ``A New Inflationary Universe Scenario: A Possible Solution Of The Horizon,
% Flatness, Homogeneity, Isotropy And Primordial Monopole Problems,''
  Phys.\ Lett.\  B {\bf 108}, 389 (1982);
  %%CITATION = PHLTA,B108,389;%%
  A.~J.~Albrecht and P.~J.~Steinhardt,
%  ``Cosmology For Grand Unified Theories With Radiatively Induced Symmetry
%  Breaking,''
  Phys.\ Rev.\ Lett.\  {\bf 48}, 1220 (1982).
  %%CITATION = PRLTA,48,1220;%%

\bibitem{linde83}
  A.~D.~Linde,
%  ``Chaotic Inflation,''
  Phys.\ Lett.\  B {\bf 129} (1983) 177.
  %%CITATION = PHLTA,B129,177;%%
  
\bibitem{irst}  
  G.~Isidori {\it et al.},
% ``Gravitational corrections to Standard Model vacuum decay,''
  Phys.\ Rev.\  D {\bf 77}, 025034 (2008).
%  [arXiv:0712.0242 [hep-ph]].
  %%CITATION = PHRVA,D77,025034;%%

\bibitem{bscd} 
  F.~L.~Bezrukov and M.~Shaposhnikov,
%  ``The Standard Model Higgs boson as the inflaton,''
  Phys.\ Lett.\  B {\bf 659}, 703 (2008).
%  [arXiv:0710.3755 [hep-th]].;
  %%CITATION = PHLTA,B659,703;%%

\bibitem{running} 
  L.~Maiani, G.~Parisi and R.~Petronzio,
%  ``Bounds On The Number And Masses Of Quarks And Leptons,''
  Nucl.\ Phys.\  B {\bf 136}, 115 (1978);
  %%CITATION = NUPHA,B136,115;%%    
  C.~D.~Froggatt and H.~B.~Nielsen,
 % ``Standard Model Criticality Prediction: Top mass 173 +/- 5 GeV and Higgs
 % mass 135 +/- 9 GeV,''
  Phys.\ Lett.\  B {\bf 368}, 96 (1996);
  %[arXiv:hep-ph/9511371].
  %%CITATION = PHLTA,B368,96;%%
  J.~Ellis {\it et al.}, 
  %J.~R.~Espinosa, G.~F.~Giudice, A.~Hoecker and A.~Riotto,
  %``The Probable Fate of the Standard Model,''
  Phys.\ Lett.\  B {\bf 679}, 369 (2009).
  %[arXiv:0906.0954 [hep-ph]].
  %%CITATION = PHLTA,B679,369;%%
  
    \bibitem{wmap5} 
  E.~Komatsu {\it et al.}  [WMAP Collaboration],
%  ``Five-Year \\
%  Wilkinson Microwave Anisotropy Probe
 % Observations: \\
 % Cosmological Interpretation,''\\
  Astrophys.\ J.\ Suppl.\  {\bf 180}, 330 (2009).
 % [arXiv:0803.0547 [astro-ph]].
  %%CITATION = APJSA,180,330;%%

  
 \bibitem{RS}
E.~Ramirez and D.~J.~Schwarz, Phys.\ Rev.\  {\bf D 80}, 023525 (2009).

\bibitem{fjse} 
  C.~Ford {\it et al.}, 
  %D.~R.~T.~Jones, P.~W.~Stephenson and M.~B.~Einhorn,
%  ``The Effective potential and the renormalization group,''
  Nucl.\ Phys.\  B {\bf 395}, 17 (1993).
  %[arXiv:hep-lat/9210033].
  %%CITATION = NUPHA,B395,17;%%
  
\end{thebibliography}
\end{document}